# Optical and electrical properties of textured sulfur hyperdoped silicon: a thermal annealing study

Ke-Fan Wang [1, 2], Shengchun Qu [2, a)], Yuanxu Wang [1, b)], Zhanguo Wang [2]

[1]*Institute for Computational Materials Science, School of Physics and Electronics, Henan University, Kaifeng 475004, P. R. China*

[2]*Key Laboratory of Semiconductor Materials Science, Institute of Semiconductors, Chinese Academy of Sciences, Beijing 100083, P. R. China*

**Abstract**: When the sulfur element is hyperdoped into crystalline silicon to a supersaturated density of $\sim 10^{20}$ cm$^{-3}$, it can enhance the sub-bandgap light absorption of silicon from 0 to 70%, with the antireflection of surface dome structures. On the other hand, the local Si: S configuration that can contribute to the strong sub-bandgap absorption is still unknown. In order to find more characteristics of this local Si:S configuration, we thermally annealed the textured sulfur hyperdoped silicon, and analyzed the changes of its optical and electrical properties. We find that the imaginary part ($\varepsilon_2$) of the complex dielectric constant of this Si:S configuration is almost constant in the wavelength range from 1250 nm to 2500 nm, and it decreases quickly when thermally annealed at the

---

a) E-mail: qsc@semi.ac.cn
b) E-mail: wangyx@henu.edu.cn

temperature range from 400℃ to 700℃. To consider the annealing process as a decomposition reaction of this local Si:S configuration, we obtain the thermal activation energy of attenuated infrared light absorption is 0.356eV for this Si:S configuration transforming from optically active state to be inactive state. Hall measurements demonstrate that thermal annealing can convert this local Si:S configuration from electrically active state to inactive state—namely, to lower the electron density and meanwhile enlarge the electron mobility. These experimental results will be very helpful to clarify the local Si:S configuration that can absorb strongly the light at the silicon sub-bandgap wavelength.

**Keywords**: sulfur hyperdoped silicon; sub-bandgap light absorption; surface texture; thermal annealing; Hall measurement;

## 1. INTRODUCTION

When the single-crystal silicon is hyperdoped by chalcogen elements (S, Se or Te) to a supersaturated density of ~$10^{20}$cm$^{-3}$ (~1% in atomic percent), its sub-bandgap light absorption can be enhanced largely.[1,2] As a result, this new type of silicon material can be potentially utilized in the fields such as high-efficiency silicon solar cells[3-6] and infrared photodiodes.[7-9] Indeed, the present silicon solar cells cannot absorb the solar light at the wavelength larger than 1100 nm, and thus lose about one-third of the total solar radiation.

At present, chalcogens hyperdoping to crystalline silicon can be made through two different methods. The first method is to structure the silicon surface using ultrafast (femo-, pico, or nanosecond) laser pulses in a chalcogen-bearing environment.[10-13] This method can produce the dense and sharp silicon spikes on the silicon surface. The surface of the spikes is covered by a thin hyperdoped layer, and the sample has a near 100% light absorption in a very wide wavelength range from ultraviolet (250 nm) to infrared (2500 nm). In spite of such an excellent light absorption, however, this kind of hyperdoped silicon is not preferable for the device fabrication. The first reason is that the sharp spikes make the good electrode contact technically difficult. The second reason is that the large amounts of surface defects formed during laser structuring can quench the photocarriers quickly. Recently, another method that also can produce hyperdoped silicon was reported.[2] It includes two steps, namely chalcogen ion implantation and nanosecond laser melting. Hyperdoped silicon formed in this way has a smooth surface, a perfect crystalline, and a comparatively low light absorption compared with the sample prepared by former method.[2,4] In general, these properties make this type of hyperdoped silicon not only suitable for the fabrication of silicon-based solar cells[4] and infrared detectors,[8,9] but also an ideal material for the study of insulator-to-metal transition in the special case of deep levels.[14,15] *So in the rest of this paper, we will limit our discussion on this new type of chalcogen hyperdoped silicon only*.

Although there have been a few optoelectronic devices that are fabricated on this new type of chalcogen hyperdoped silicon,[4,8,9,16] their performance at the silicon sub-bandgap wavelength is far from satisfactory, such as the limited photoresponse wavelength,[8,9] the low quantum efficiency (less than $10^{-5}$ and $10^{-7}$ at wavelength 1310 nm and 1550 nm, respectively),[16] in contrast to their strong and broad infrared absorption. In order to improve the performance of these optoelectronic devices, the exact local Si:S configuration that is responsible for the strong infrared absorption should be clarified firstly. Unfortunately, this Si:S configuration still cannot be confirmed, although there are a few theoretical calculations suggesting some possible configurations.[17,18] Mo et al [17] performed the ab initio calculations of sulfur configurations and energetics in crystalline silicon and found that sulfur prefers to occupy the substitution site that will produce deep levels in the silicon forbidden band and thus lead to the sub-bandgap absorption. They also found that upon annealing, the sulfur atoms in the hyperdoped silicon tend to form the dimmers, and the dimmers can remain the infrared absorbing, which is contrary to the experimental findings.[2] More recently, Shao et al.[18] also theoretically investigated sixteen typical Si:S configurations, and found that among them, eight quasi-substitutional and four interstitial metastable configurations can introduce impurity levels into the silicon forbidden band, and after thermal annealing, these Si:S configurations transform to be three more stable two-coordinated interstitial configurations, which cannot introduce any impurity level into the silicon forbidden band. Their calculations can explain many experimental results reasonably, but they still didn't elucidate which local Si:S configuration can induce the strong sub-bandgap absorption, and this configuration will transform to be which more stable configurations when thermally annealed.

In this study, firstly we report that the sulfur hyperdoped silicon was prepared following a new process that includes surface texture, sulfur ion implantation and nanosecond laser melting. Then we thermally annealed the obtained hyperdoped silicon at different temperatures, and analyzed the changes of its optical and electrical properties, in order to find more characteristics about the local Si:S configuration that can induce the strong infrared absorption.

## 2. EXPERIMENTAL DETAILS

390-μm thick, B doped p-type Si wafers (1~10 Ω·cm, double-side polished, Czochralski-grown, (100) orientation) were surface textured by a deionized water solution containing sodium-hydroxide and isopropanol.[19] Then the textured Si wafers were ion implanted at 200 keV with [32]S+ at 7° off-normal to prevent channeling effect, with a dose of $1\times10^{16}$ cm[-2]. Finally, the ion implanted samples were laser melted by a spatially homogenized pulsed KrF excimer laser beam (248 nm, 20 ns, rectangle in size 3 mm×1 mm). The average point on the surface was irradiated by four pulses. Another Si wafer, treated by the same process except for the surface texture, served as a control sample. The hyperdoped silicon wafer was cleaved into many 1cm × 1cm squares for thermal annealing that was carried out in a horizontal tube furnace, flushed continuously with a high purity $N_2$ gas.

Sample morphology was observed by a JSM-6301F scanning electron microscope (SEM) system. The total hemispherical (specular and diffuse) reflectance (R) and transmittance (T) of the samples were measured using an AvaSpec-2048 UV-VIS-NIR spectrometer over the wavelength range from 400 nm to 2500 nm. The spectrometer was equipped with an integrating sphere detector. The total absorptance (A = 1- T- R) of the samples was determined from the directly measured T and R. The instruments were calibrated by sets of Labsphere Diffuse Reflectance Standards. The Hall coefficient, carrier density, carrier mobility and conductivity of the hyperdoped silicon sample were measured at room temperature by a Hall measurement using the *van der pauw* method.

## 3. RESULTS AND DISCUSSION

### 3.1 Preparation of sulfur hyperdoped silicon samples

After the silicon (001) wafer was chemically textured, its surface was covered with dense pyramids, as shown in Fig. 1(a). The formation of the pyramid can be attributed to the anisotropy etching rate between Si ⟨001⟩ and ⟨111⟩ directions.[19] Subsequently, the textured silicon wafer was sulfur ion implanted. The depth profile of sulfur ion inside the textured silicon wafer can be simulated by SRIM software,[20] as shown in Fig. 1(c), in view of the angel 54.7° between the incident ion beam and the normal direction of the Si (111) side faces. We can see from Fig. 1(c) that the sulfur ions can reach as deep as 350 nm beneath the textured surface, and the peak doping density is about $5 \times 10^{20}$ cm$^{-3}$, corresponding to an atom percent of ~1% in silicon. Then the implanted silicon sample was laser melted at a fluence of 1.0 J/cm$^2$. The surface morphology of the sample after laser melting is shown in Fig. 1 (b). One can find that the small pyramids change to be small rounded domes while the large pyramids to be big domes, still with four symmetrical ridges. The reason for this change of morphology is that laser pulses can melt the surface of the pyramid to be liquid, which in turn can round the angular pyramids via surface tension. From the visible change of pyramid surface, we can safely conclude that the whole ion implanted layer (~350 nm) was melted completely. Previous transmission electron microscope (TEM) observation had confirmed that the implantation-induced lattice damages can be restored and recrystallized completely after nanosecond laser melting,[4] as is also consistent with its *constant* characteristic of infrared absorption (See the curve "Tex-HD Si" in Fig. 2(a)).[2,22]

Meanwhile, we chose another same Si (001) wafer as a control sample. It was treated following the same process as before but except for the surface texture. Its original surface morphology is shown in Fig. 1 (d). After it was ion implanted, its smooth surface did not change. Fig. 1(f) presents that the sulfur ion can reach ~430 nm beneath the flat surface. In order to melt this ion implanted layer, we increased the laser fluence from 1.0 to 1.8 J/cm$^2$, according to a previous literature.[21] After the laser melting, the sample surface remains smooth (See Fig. 1(e)) although with a higher laser fluence. To explain this point, we measured the reflectances of the textured and nontextured silicon samples at the work wavelength 248 nm of KrF laser (not shown here), and they are ~4% and ~47%, respectively. This low reflectance of textured silicon samples at 248 nm is considered to relate to the morphology of the pyramid, the wavelength of incident light, and the optical property of silicon after sulfur ion implantation. That is to say, on the textured surface almost all the laser radiation can enter into the wafer surface while on the nontextured surface only one-half can enter. So the temperature on the textured surface should be much higher than that on the nontextured surface. This is the reason why the textured surface can melt remarkably while the nontextured surface cannot.

Fig. 2 shows the optical properties of the nontextured (See the curve "NonTex-HD Si") and textured (See the curve "Tex-HD Si") hyperdoped silicon samples, as well as the pristine Si (001) wafer (See the curve "Untreated Si"). We find that the optical properties of the nontextured hyperdoped silicon sample in the infrared wavelength are similar to that previously reported.[23] Comparing the curve "NonTex-HD Si" with "Tex-HD Si", we find that the absorptance of the former is 15% less than that of the latter in the visible wavelength, and 40% less in the infrared wavelength. In the visible wavelength, their transmittances are both 0, but the textured silicon has a 15% lower reflectance than that of the nontextured silicon, and so the former has a 15% higher absorptance. In the sub-bandgap wavelength (> 1200 nm), the textured hyperdoped silicon has a 10% fewer reflectance and a 30% less transmittance than those of the nontextured hyperdoped silicon, and so the former has a 40% larger absorptance, as shown in Fig. 2(a). We have

known that for the nontextured and textured hyperdoped silicon wafers, the ion implanted

layer has been melted completely by the pulsed nanosecond laser. However, on the

textured surface the melted hyperdoped silicon will flow down from the side faces of the

pyramids, and meanwhile it also can round the angular pyramids via surface tension, as

indicated by Figs. 1(a) and 1(b). Such behaviors will induce a mass transport of S

dopants and thus a larger fluctuation of S distribution in the textured sample surface than

that in the nontextured sample surface. But this fluctuation should not affect the optical

absorption significantly because that the absorption is only related to the total S

concentration,[22] which does not change statistically. On the other hand, we find one obvious difference between the nontextured and textured silicon samples is that, the former has a smooth surface while the latter owns a dome-covered surface. The rough silicon surface can not only reduce the reflection of the incident light, but also lengthen the light effective path inside the hyperdoped silicon through multiple interface reflection, and finally reduce the transmittance. In one word, surface texture can enhance the infrared absorption of hyperdoped silicon from 30% to 70% and should be a necessary step in the fabrication of solar cells and infrared photodetectors that is based on sulfur hyperdoped silicon.

Another noteworthy point is that in comparison with the pristine Si wafer (See the

curve "Untreated Si" in Fig. 2(a)), the sulfur hyperdoping (See the curve "NonTex-HD

Si") can enhance the silicon sub-bandgap absorptance from 0 to 30%. It has been well accepted that the impurity levels, filled in the band gap, produce this broad and featureless absorption,[2,15] although the local Si:S configuration that can introduce the impurity levels into the band gap is still unclear.

## 3.2 Optical properties of the annealed hyperdoped samples

In order to collect more characteristics of this local Si:S configuration, we chose the textured hyperdoped silicon samples, melted with a laser fluence of 1.0 J/cm$^2$, and performed thermal annealing on them at different temperatures for 30 minutes. Optical properties of these annealed samples are shown in Fig. 3. In the visible wavelength (400 nm~1000 nm), thermal annealing does not induce any obvious changes in their optical spectra. In the infrared wavelength (>1200 nm), however, thermal annealing at temperatures higher than 300℃ increases their reflectances and transmittances obviously, and thus reduce their absorptances. This decrease of infrared absorption after thermal annealing has also been observed by other researchers.[2,24] But the infrared absorption of our textured hyperdoped samples is more persistent to thermal annealing than that of the nontextured ones. For instance, after thermal annealing at 600℃ for 30 minutes, our textured samples can preserve ~44% of the infrared absorption of the unannealed samples while the nontextured samples of Kim et al.[2] only keeps ~26%.

Next, we will investigate the effect of annealing temperature on the optical constants of the textured hyperdoped layer at 2000 nm. Due to the multiple reflection taking place at the rough interface between the air and the textured hyperdoped layer, and at the internal interface between the textured hyperdoped layer and the Si substrate, as well as at the flat interface between the back surface of Si substrate and the air, it is very difficult to perform a optical analysis. In order to simplify this optical model, firstly we neglect the reflection at the internal interface between the hyperdoped layer and the Si substrate, which is considered to be reasonable;[25,26] secondly, the rough textured hyperdoped structure is equated to be a flat one. This treatment of equating will not affect the conclusion of optical analysis because that here we only care about the variation trend, rather than the accurate values, of the optical constants of hyperdoped samples after thermal annealing, and also because that the different annealed hyperdoped samples own the same surface morphology, sample structure and optical measurement condition. After the above two simplications, the transmittance and reflectance of the hyperdoped sample can be given by,[25]

$$T = \frac{(1-r_{hd})(1-r_{Si})\exp(-\alpha_{hd}d_{hd})\exp(-\alpha_{Si}d_{Si})}{1-r_{hd}r_{Si}\exp(-2\alpha_{hd}d_{hd})\exp(-2\alpha_{Si}d_{Si})} \qquad (1)$$

$$R = r_{hd} + \frac{\left(1-r_{hd}\right)^2 r_{Si}\exp(-\alpha_{hd}d_{hd})\exp(-2\alpha_{Si}d_{Si})}{1-r_{hd}r_{Si}\exp(-2\alpha_{hd}d_{hd})\exp(-2\alpha_{Si}d_{Si})} \qquad (2)$$

where

$r_{hd}$ is the reflectivity coefficient of the interface between air and the hyperdoped layer,

$r_{Si}$ is the reflectivity coefficient of the interface between air and the Si substrate,

$\alpha_{hd}$ is the absorption coefficient of the hyperdoped layer,

$\alpha_{Si}$ is the absorption coefficient of the Si substrate,

$d_{hd}$ is the thickness of the hyperdoped layer, and

$d_{Si}$ is the thickness of the Si substrate.

The functional relations between the absorption coefficient $\alpha$ and extinction coefficient $k$ is given by,[27]

$$\alpha = \frac{4\pi\nu k}{c} \tag{3}$$

in which $\nu$ and $c$ denote the frequency of incident light and the speed of light in the vacuum, respectively. For normal incidence, the reflectivity coefficient $r$ is given by,[27]

$$r = \frac{(n-1)^2 + k^2}{(n+1)^2 + k^2} \tag{4}$$

where $n$ is the refractive index and $k$ is the extinction coefficient.

According to the values of $n$ (3.449) and $k$ (~0) for crystalline silicon at 2000 nm,[28] we obtain the $r_{Si}$ and $\alpha_{Si}$ are 0.5042 and 0, respectively. Using these specific values to replace the $r_{Si}$ and $\alpha_{Si}$ in Equation (1) and (2), and meanwhile to substitute the $T$ and $R$ by the measured transmittance and reflectance, we can numerically solve the Equation (1) and (2) and obtain a series of the values of $r_{hd}$ and $\alpha_{hd}d_{hd}$ for the annealed hyperdoped layers at different annealing temperatures.

We replace the absorption coefficient $\alpha$ in Equation (3) by the values of $\alpha_{hd}d_{hd}$, and obtain the values of $k_{hd}d_{hd}$ for different annealed hyperdoped samples. Here the symbol $k_{hd}$ denotes the extinction coefficient of the sulfur hyperdoped layer. Since the value of $d_{hd}$ is same with each other among the different annealed samples, we divide these $k_{hd}d_{hd}$ by that of the unannealed sample in order to eliminate the effect of $d_{hd}$ and thus obtain the normalized $k_{hd}$, as shown in Fig. 4(a).

For the sulfur hyperdoped layer, the $k^2$ is usually very small (~0.1 from Fig. 4 in Ref. 22), so the Equation (4) approximates to be,

$$r \approx \frac{(n-1)^2}{(n+1)^2} \qquad (5)$$

To replace the $r$ in Equation (5) by the obtained $r_{hd}$ above, we can obtain the values of $n_{hd}$ for the different annealed hyperdoped layers. Similar as that doing for $k_{hd}$, we can obtain the normalized $n_{hd}$, as shown in Fig. 4(b).

The imaginary part ($\varepsilon_2$) of the complex dielectric constant can be expressed in term of refractive index $n$ and extinction coefficient $k$ as,[29]

$$\varepsilon_2 = 2nk \qquad (6)$$

Substituting the $n$ and $k$ in Equation (6) by the normalized $n_{hd}$ and $k_{hd}$ obtained above, we can achieve the normalized $\varepsilon_2$ of the different annealed hyperdoped silicon, as shown in Fig. 4(c).

We can find from Fig. 4 that as the annealing temperature increases, the refractive index $n$ almost keeps constant, with a little fluctuation of ±10%, whereas the extinction coefficient $k$ and the imagery part ($\varepsilon_2$) of the complex dielectric index decrease simultaneously. Especially, the $\varepsilon_2$ decreases quickly when the annealing temperature increases from 400 ℃ to 700 ℃. The decrease of $k$ and $\varepsilon_2$ is due to the decrease of light absorption after thermal annealing. Except for the wavelength of 2000 nm, we can deduce from Fig. 3 that the value of $\varepsilon_2$ is almost constant in the wavelength range from 1250 nm to 2500 nm. Since the $\varepsilon_2$ can be obtained from the electronic structure calculation by first-principles method,[18] such characteristics of $\varepsilon_2\square$ can help to distinguish the objective Si:S configuration from others.

We also plot the product ($\alpha_{hd}d_{hd}$) of absorption coefficient $\alpha_{hd}$ and effective thickness $d_{hd}$ at 2000 nm as a function of $1/kT$, as shown in Fig. 5. It is noteworthy that here $k$ is the Boltzmann constant and $T$ is the annealing temperature (And with a unit in Kelvin). At the annealing temperature of 200 ℃, the absorption of the hyperdoped silicon does not change. However, the annealing at 300 ℃ begins to wipe the light absorption. At the higher annealing temperatures, the absorption decreases exponentially. This exponent shape inspires us that the attenuation of infrared absorption during thermal annealing may be considered as a process that the optically active state (denoted by "A") of the objective Si:S configuration decomposes to be the optically inactive state (denoted by "B"). The chemical reaction equation can be written as follows,

$$A \xrightarrow{\quad K \quad} B \tag{7}$$

Suppose that the concentration of "A" at the moment t, $C_t$, obeys such a relation,

$$-\frac{dC_t}{dt} = KC_t \tag{8}$$

in which $K$ is the reaction constant. To solve this differential equation, we obtain,

$$K = \frac{1}{t} \ln \frac{C_0}{C_t}$$

(9)

in which $C_0$ denotes the concentration of "A" at the moment of beginning. Suppose that $C_t \propto \alpha_t$, and $C_0 \propto \alpha_0$, where $\alpha_\tau$ and $\alpha_0$ are the absorption coefficients at the moment $t$ and at the moment of beginning, respectively. Then Equation (9) can be written as follows,

$$K = \frac{1}{t} \ln \frac{\alpha_0}{\alpha_t}$$

(10)

Assume that the decomposition reaction is a thermal activation process, then

$$K = A \exp(-E_a / kT)$$

(11)

To combine Equation (10) and (11), we obtain,

$$\alpha_t = \alpha_0 \exp(-At \exp(-Ea / kT))$$

(12)

We multiply both sides of Equation (12) by the effective thickness $d$. This equation converts to be,

$$\alpha_t d = \alpha_0 d \exp(-At \exp(-Ea / kT))$$

(13)

We have obtained the product $\alpha_{hd} d_{hd}$ above and denoted them by "■" in Fig. 5. Using Equation (13) to fit these $\alpha_{hd} d_{hd}$ points, we obtain a very good fitting curve (See the solid line in Fig. 5), which indicates that our supposition is very reasonable. According to the fitting result, we find that the thermal activation energy $E_a$ is about $0.356 \pm 0.018$ eV. The $E_a$ is the height of energy barrier that prevents the optically active state "A" from transforming to be optically inactive state "B". This explicit value of $E_a$ is very useful to clarify the objective Si:S configuration that we are searching if combined with the transition state calculation.[30]

## 3.3 Electrical properties of the annealed hyperdoped samples

The electrical property of the textured hyperdoped silicon sample is also important to the future application. So we carried out the room-temperature Hall measurements on them. In addition, since the carrier density is on the order of $10^{19}$cm$^{-3}$, as will be shown in Fig. 6(b), the hyperdoped sample is actually an n$^+$/p junction with a depth of ~350 nm beneath the curved surface of the dome structures (See Fig. 1(c)), which will naturally isolate the electron transporting from the underlying p-type silicon substrate.[4] As a result, the obtained carrier property is that limited inside the surface hyperdoped layer. On the other hand, when the textured hyperdoped sample is for Hall measurement, the curved surface makes the direction of electron moving is not vertical with the direction of the imposed magnetic field. As a result, the Hall coefficient $R$ is,

$$R = \pm \frac{\sin\theta}{nq} \qquad (14)$$

where $\theta$ is the angle between the moving direction of the electron and the direction of the magnetic field, $n$ is the electron density and $q$ is the charge of electron. Since the $sin\theta$ is always less than 1, the absolute value of $R$ is less than $1/nq$. And because that the $\theta$ changes frequently during electron moving, it is difficult to estimate the value of $sin\theta$. If we follow the usual way to obtain the carrier density $n$, i.e. let $n$ equate to $1/qR$, then we will obtain a larger $n$ than the true $n$, and thus a smaller carrier mobility $\mu$ than the true $\mu$.

That is to say, we will usually obtain a larger $n$ and a smaller $\mu$ for textured samples than that for nontextured samples, which is indeed observed in our unannealed hyperdoped samples. Since the $sin\theta$ is statistically same for different annealed samples, the correction to carrier density $n$ and mobility $\mu$ doesn't change their variation trend with the annealing temperatures. For simplicity, here we do not normalize these electrical parameters again.

Then we begin to investigate the effect of thermal annealing on the electrical properties of textured hyperdoped silicon samples, with a laser melting fluence of 1.0 J/cm$^2$. Fig. 6(a) shows that Hall coefficients of all samples are negative, which indicates that the dominant carriers in these hyperdoped silicon layers are electrons (n-type conductivity). Since the original silicon substrate is doped by B atoms (p-type conductivity) in the density of $10^{15}$~$10^{16}$cm$^{-3}$, we believe that the hyperdoped sulfur atoms can release much more electrons than that of the original holes.

Fig. 6(b) reveals that as the annealing temperature increases, the carrier density decreases slowly from $5.25\times10^{19}$cm$^{-3}$ for the sample unannealed to $8.9\times10^{18}$cm$^{-3}$ for the sample annealed at 700℃. Fig. 6(c) presents that the carrier mobility increases quickly at the annealing temperatures from 300℃ to 600℃, and then decreases a little at higher temperatures. This increase of carrier density and decrease of carrier mobility when thermally annealed were also observed by other researchers.[4] We have known from Section 3.2 that thermal annealing enables the optically active state "A" to decompose to be optically inactive state "B". Following this consideration, we deduce from Fig. 6(b) that state "A" can release electrons to the conduction band (CB) at room temperature while state "B" cannot. That is to say, thermal annealing renders the electrically active state "A" to be inactive state "B". Interestingly, the increase trend of carrier mobility is generally synchronous with the decrease trend of carrier density, as shown in Fig. 6(b). The reason may be that the scattering of active state "A" (more ions) is stronger than that of inactive state "B" (less ions). Their little asynchrony at annealing temperature 700℃ is probably due to the additional neutral scattering. This situation is very possible. For example, if some neutral impurity atoms locate in the transporting path of the mobile electrons, it will raise an additional scattering and then reduce the carrier mobility. Fig. 6(d) represents the conductivity of the hyperdoped sample varies with the annealing temperature. The conductivity is a product of the carrier density and carrier mobility. For the unannealed sample, it owns a good conductivity, mainly due to the very high carrier density although with a low carrier mobility. When thermally annealed, the carrier density of the hyperdoped silicon decreases while the carrier mobility increases, and generally the conductivity drops. But when the sample is annealed at 800℃, its conductivity increases again because of the reincreasing of carrier density. We can also find from Fig. 5 that the annealing at 800℃ begins to enhance the infrared absorption again. The changes of optical and electrical properties at this high annealing temperature indicate that the state "B" of the local Si:S configuration may begin to undergo another configuration transition. The reactivation of sub-bandgap absorption of sulfur hyperdoped silicon when thermally annealed at very high temperatures is also observed by Newman et al.[31]

We notice that even at a high annealing temperature of 700℃, there is still a low absorptance of 20% (Fig. 3(a)) and a n-type conductivity (Fig. 6(a)). This part of the contribution should come from the substitution site of S atoms, which is the most thermally stable configuration[5,18] and are both optically and electrically active.[4]

## 4. CONCLUSION

In summary, we have successfully prepared the sulfur hyperdoped silicon through a new three-step process, which includes surface chemical texture, sulfur ion implantation and nanosecond laser melting. The formed hyperdoped silicon has a very high sub-bandgap light absorption of 70%, due to the antireflection of surface dome structures. The imaginary part ($\varepsilon_2$) of the complex dielectric constant of the local Si:S configuration, which can induce a strong sub-bandgap absorption, decreases quickly in the annealing temperature range from 400℃ to 700℃. The thermal activation energy of this local Si:S configuration for the transforming from optically active state to inactive state is about 0.356eV. With the increase of annealing temperature, the decrease of carrier density in the hyperdoped silicon is synchronous with the increase of carrier mobility. These specific characteristics, if combined with the transition state calculation, can exclusively clarify the local Si:S configuration that can lead to the strong sub-bandgap light absorption in sulfur hyperdoped silicon.

## ACKNOWLEDGEMENTS

The author KFW thanks Hezhu Shao of Ningbo Institute of Materials Technology & Engineering, Chinese Academy of Sciences, for helpful discussions. This work was supported by the NSFC (Nos. 61204002, 61076009, 51371076), and by the National Basic Research Program of China (Nos. 2012CB934200).

## FIGURES AND CAPTIONS

**FIG. 1.** The left column includes the SEM images of Si(001) wafer surface (a) after chemical texture, (b) after laser melting, as well as the depth profile of sulfur ions (c); the right column contains the SEM images of the control wafer surface (d) original, (e) after laser melting, as well as the depth profile of sulfur ions (f).

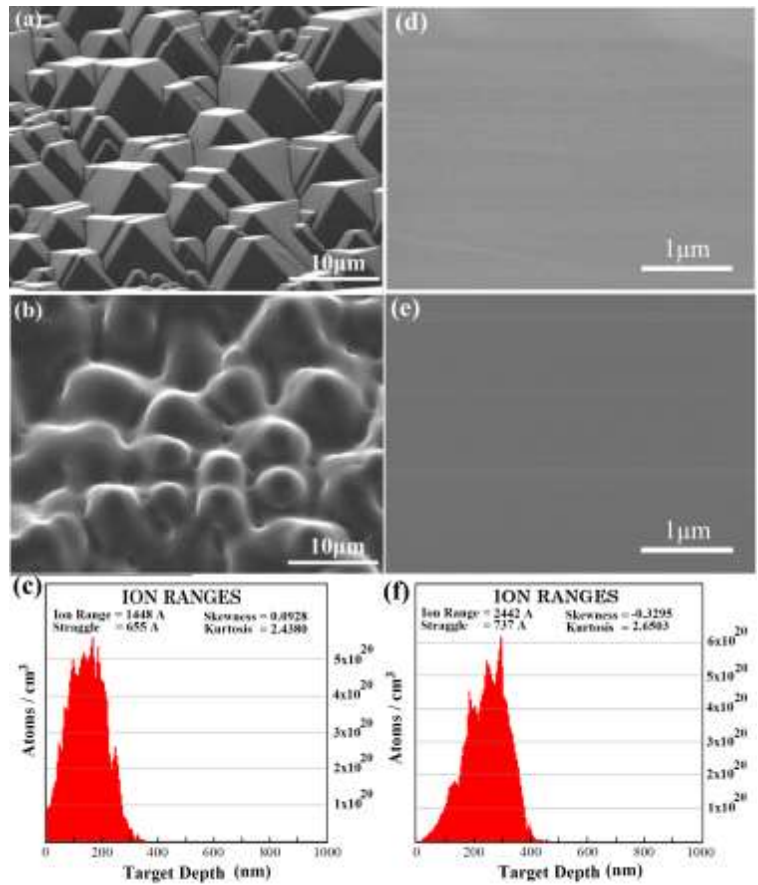

**FIG. 2.** Absorptance (a), reflectance (b) and transmittance (c) of nontextured and textured hyperdoped silicon samples as well as the pristine silicon wafer.

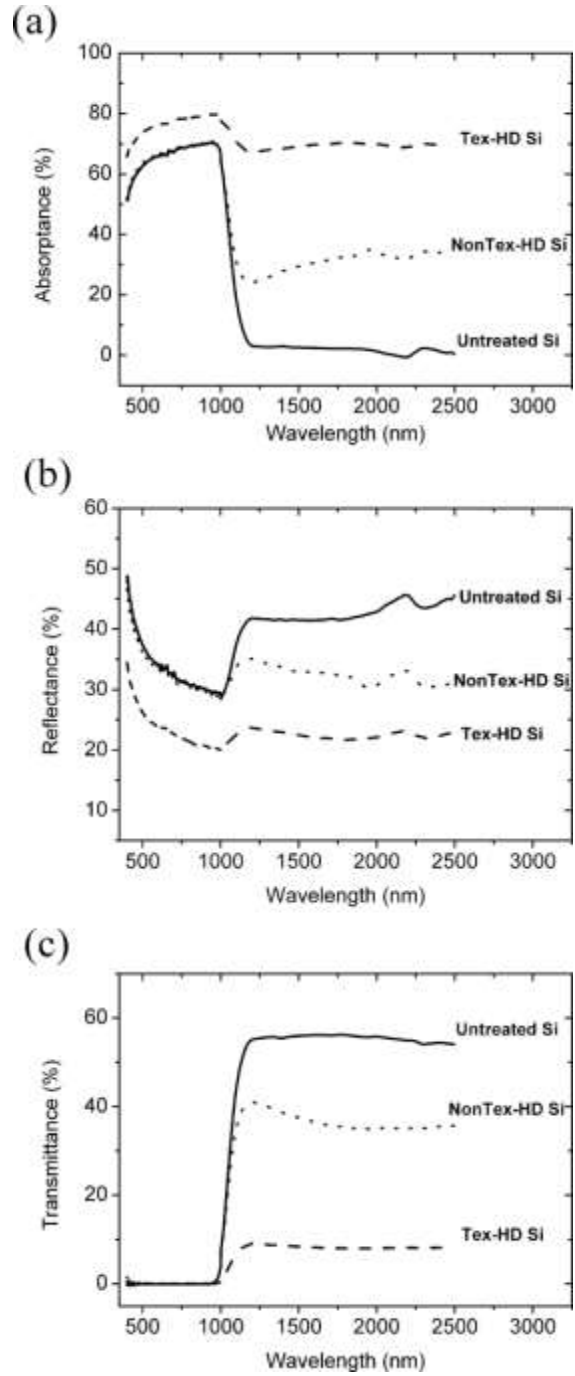

**FIG. 3.** Absorptance (a), reflectance (b) and transmittance (c) of the annealed hyperdoped silicon. The mark "HD-Si" denotes the sample before thermal annealing, and the "A-200" denotes the sample thermally annealed at 200 ℃, and so on.

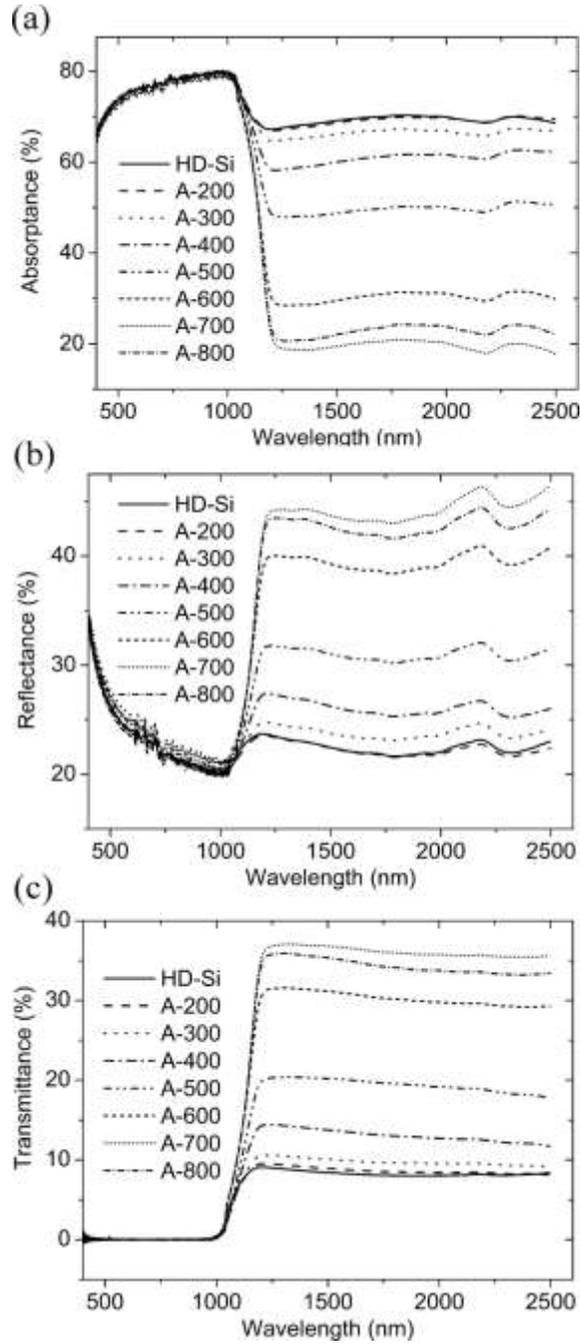

**FIG. 4.** The plots of normalized (a) extinction coefficient $k_{hd}$, (b) refractive index $n_{hd}$ and (c) the imaginary part ($\varepsilon_2$) of the complex dielectric constant of the hyperdoped silicon as a function of annealing temperatures.

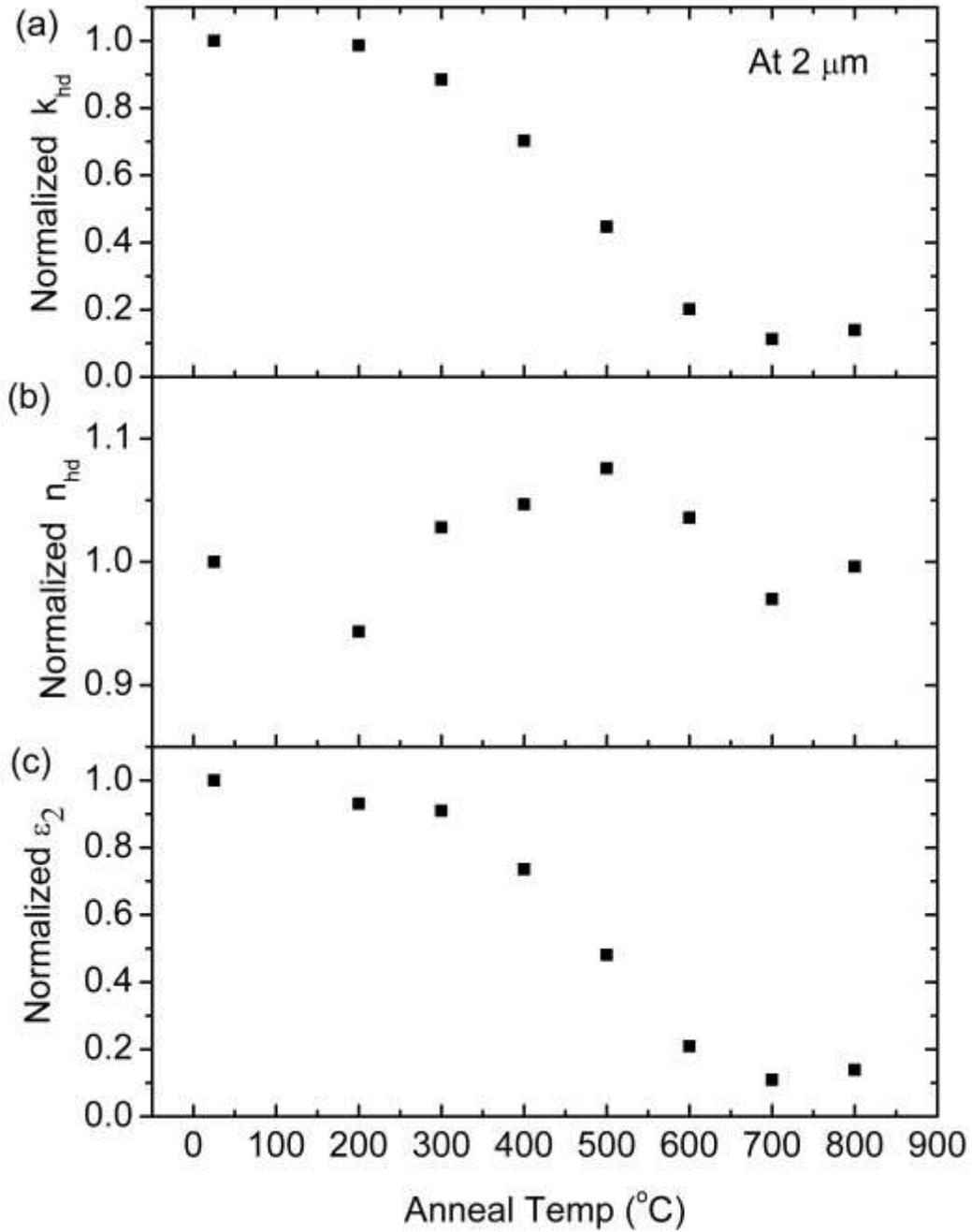

**FIG. 5.** The plot of the product $\alpha_{hd}d_{hd}$ of the annealed hyperdoped silicon at 2000 nm (denoted by "∎") as a function of $1/kT$. Here $\alpha_{hd}$, $d_{hd}$, $k$, and $T$ represent the absorption coefficient, effective thickness, Boltzmann constant, and annealing temperature (in Kelvin), respectively. The solid line is the fitting result in the annealing temperature range from 25°C to 700°C, while the leftmost "∎" of 800°C is excluded from the fitting because that it indicates a new configuration transition may begin.

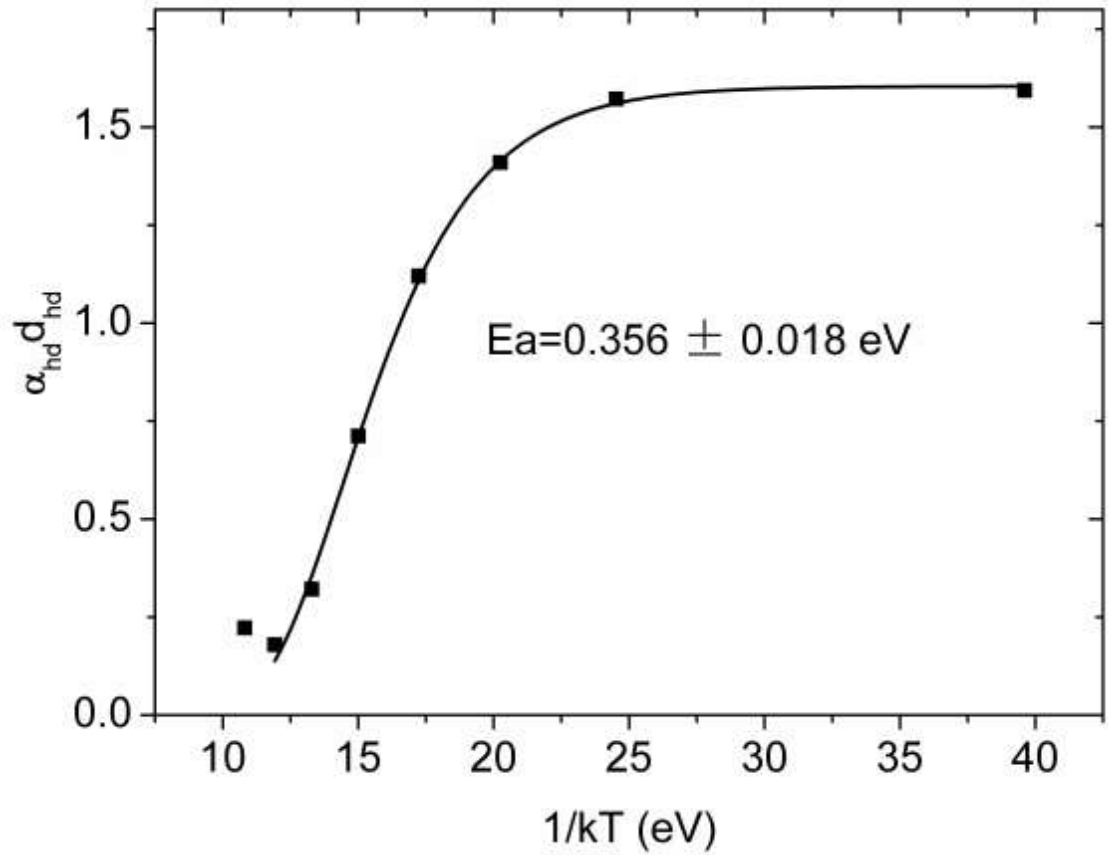

**FIG. 6.** The plots of Hall coefficients (a), carrier densities (b), carrier mobilities (c) and conductivities (d) of hyperdoped silicon as a function of annealing temperature.

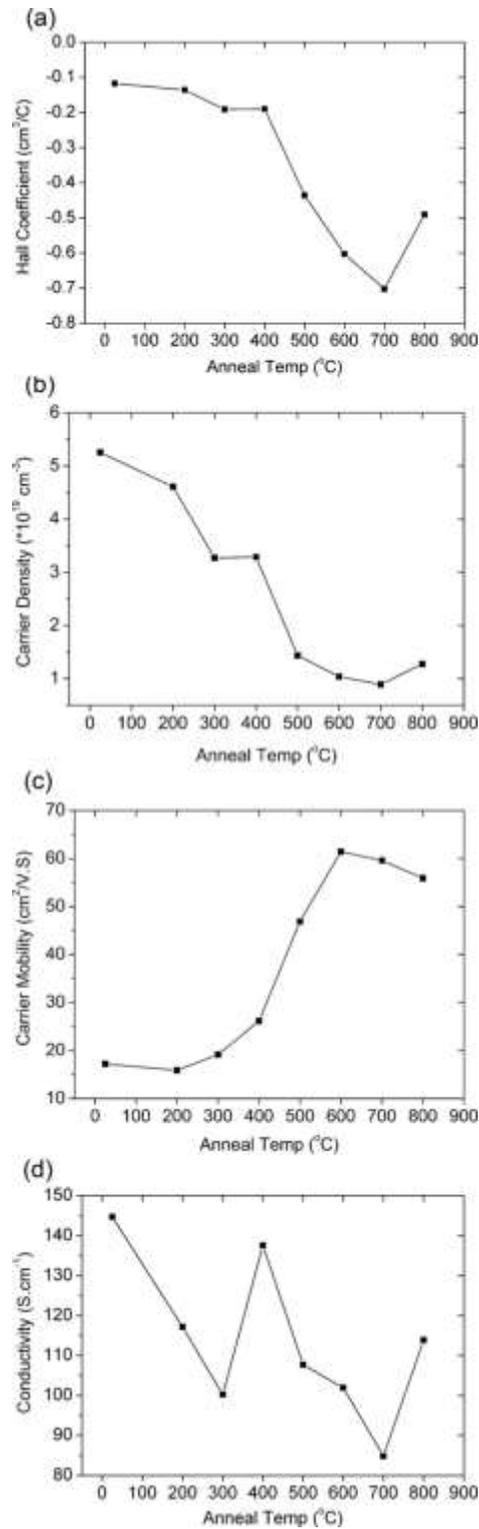

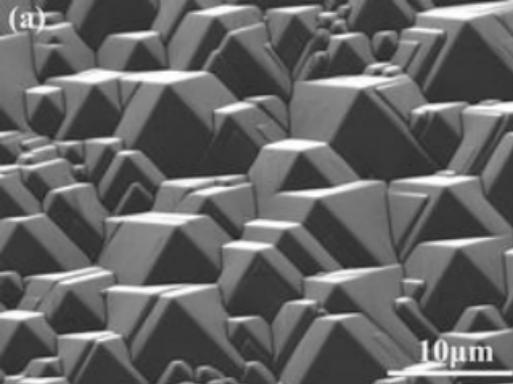

(a)

10μm

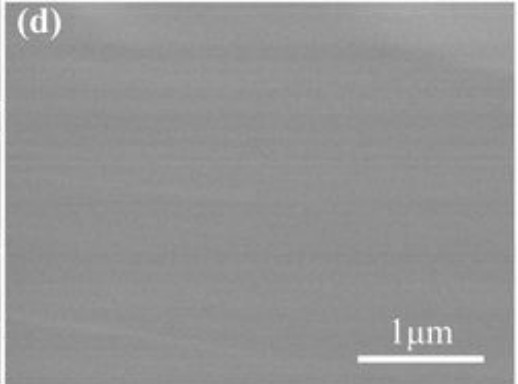

(d)

1μm

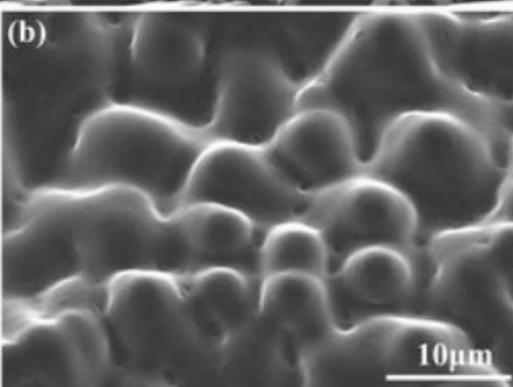

(b)

10μm

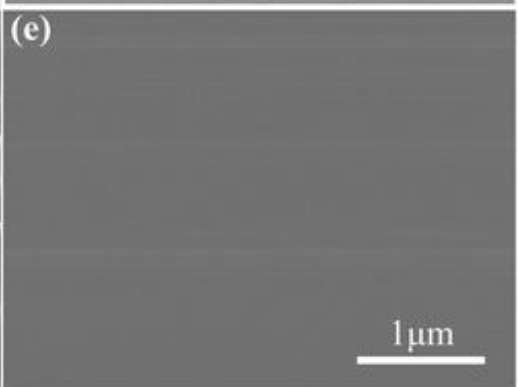

(e)

1μm

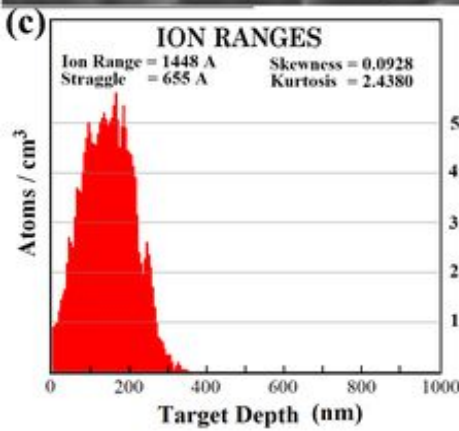

(c)

**ION RANGES**

Ion Range = 1448 A    Skewness = 0.0928
Straggle   = 655 A     Kurtosis = 2.4380

Atoms / cm³

5x10²⁰
4x10²⁰
3x10²⁰
2x10²⁰
1x10²⁰

0   200   400   600   800   1000
Target Depth (nm)

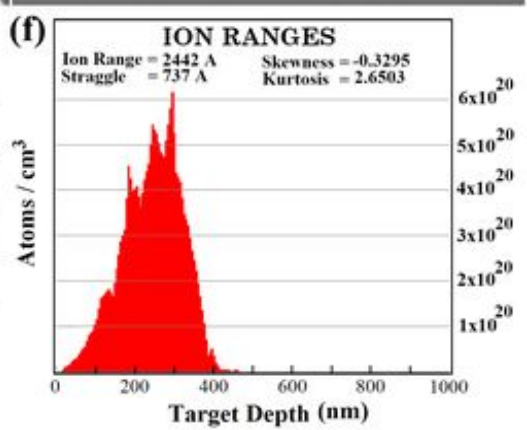

(f)

**ION RANGES**

Ion Range = 2442 A    Skewness = -0.3295
Straggle   = 737 A     Kurtosis = 2.6503

Atoms / cm³

6x10²⁰
5x10²⁰
4x10²⁰
3x10²⁰
2x10²⁰
1x10²⁰

0   200   400   600   800   1000
Target Depth (nm)

(a)

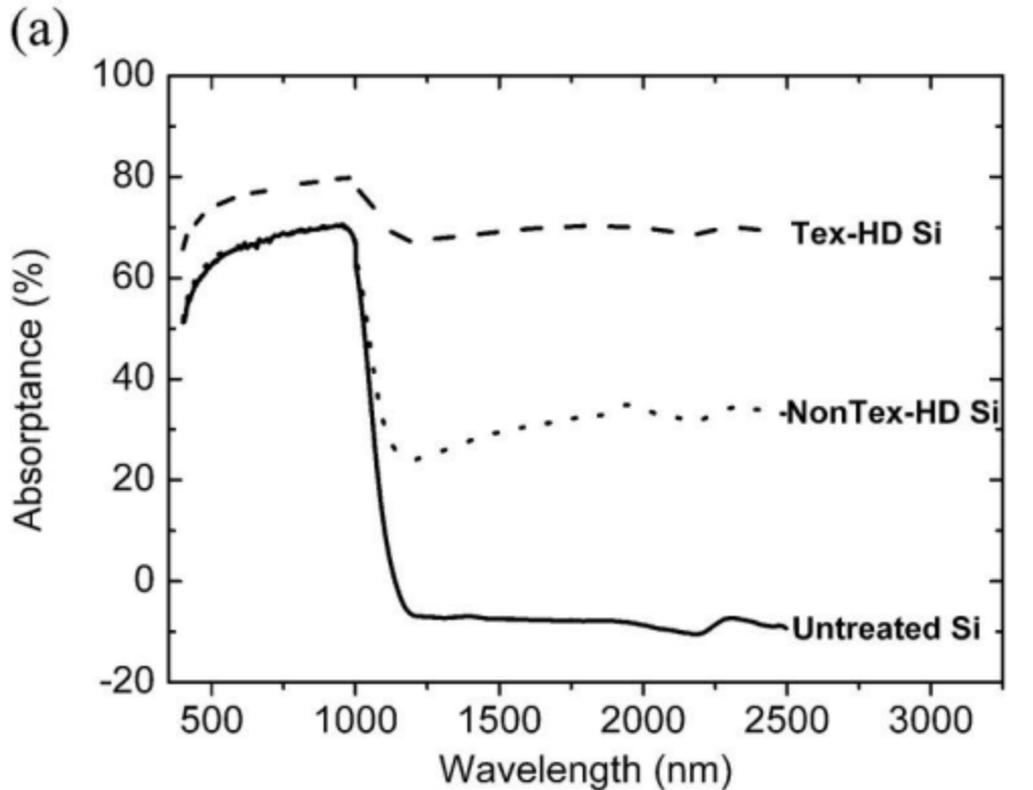

(b)

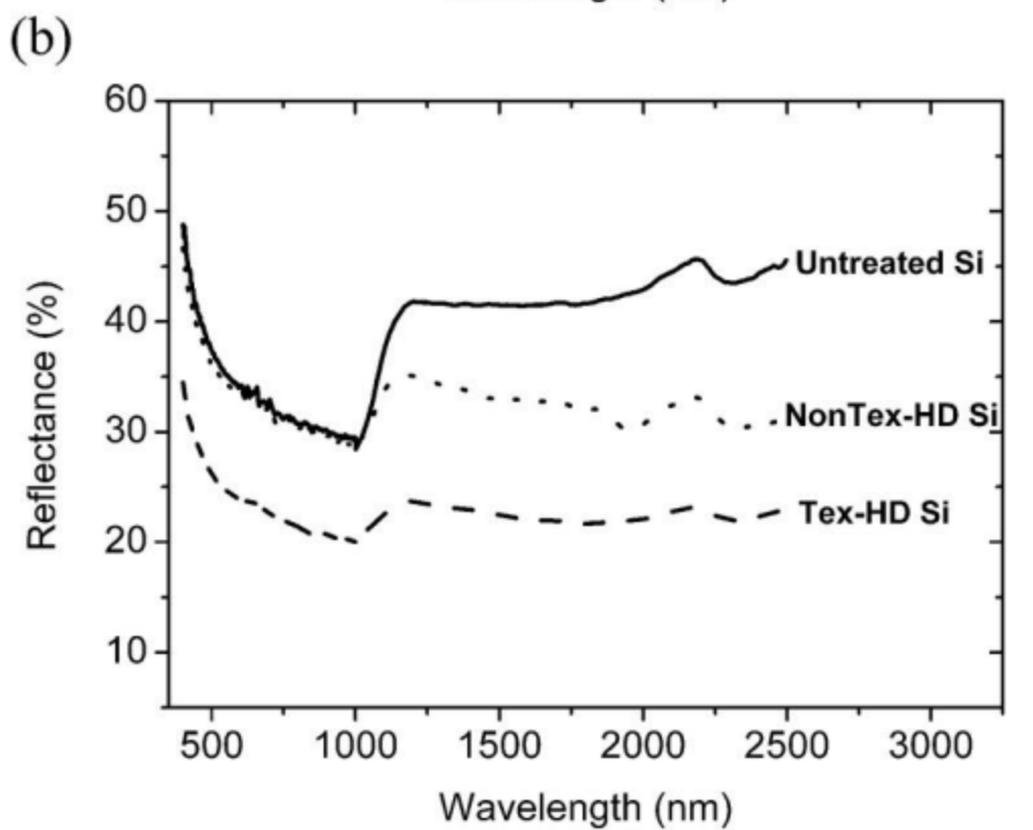

(c)

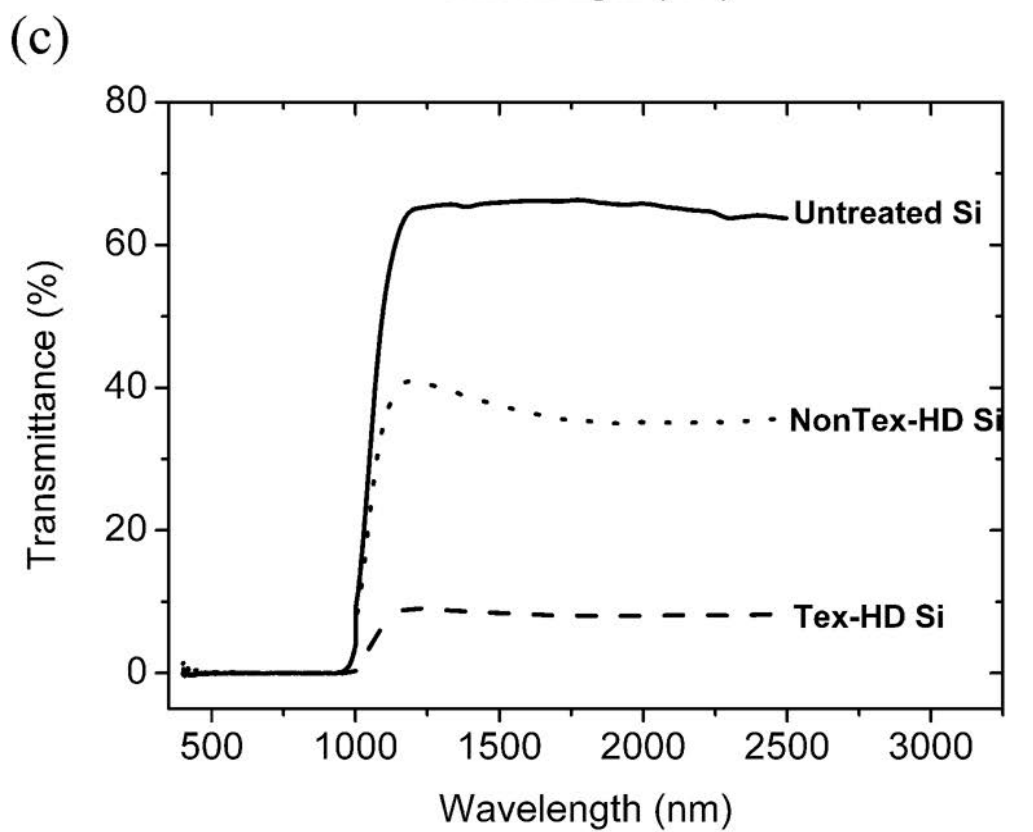

(a)

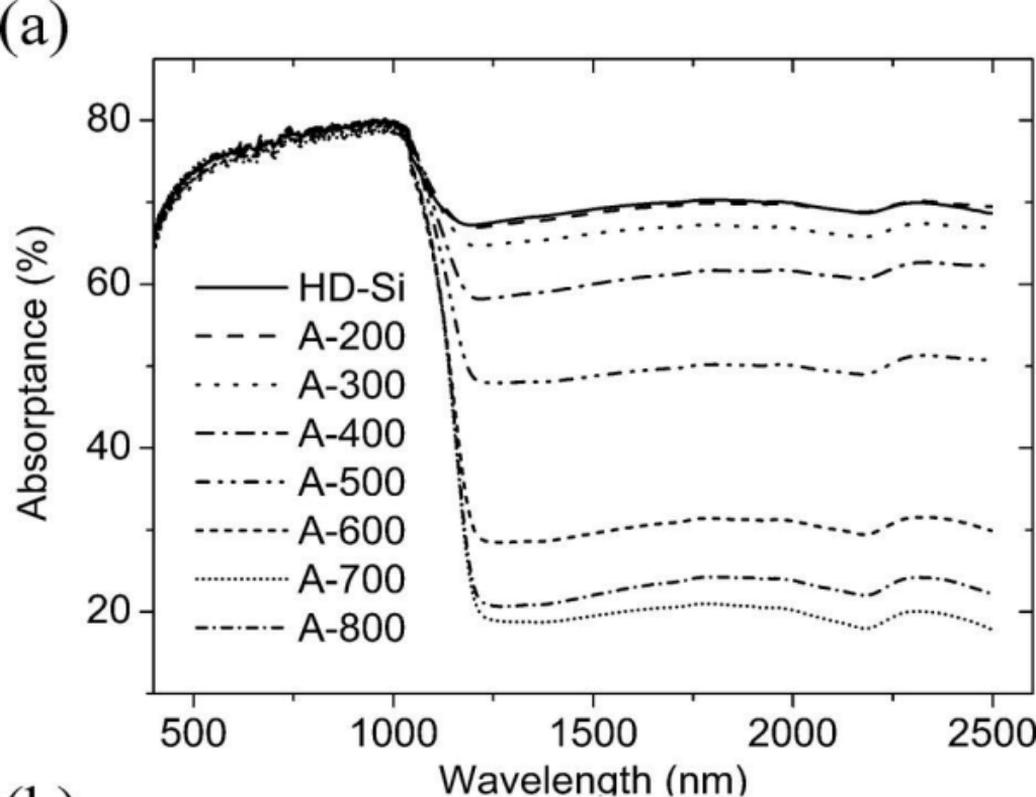

(b)

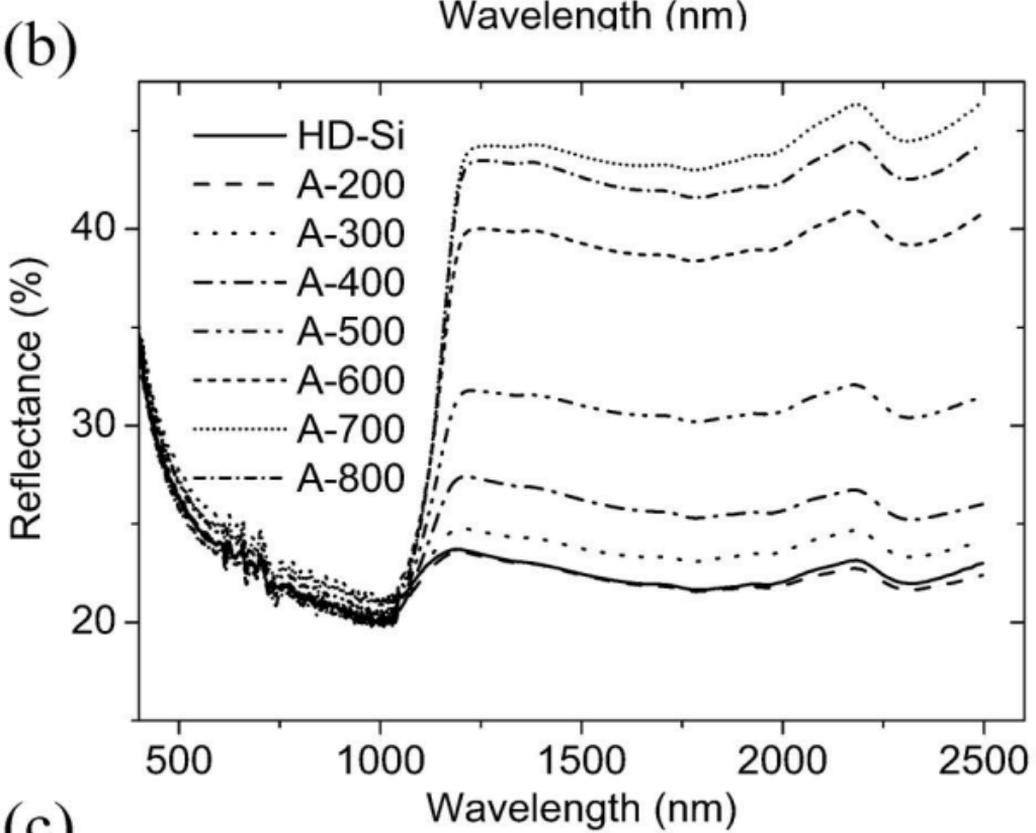

(c)

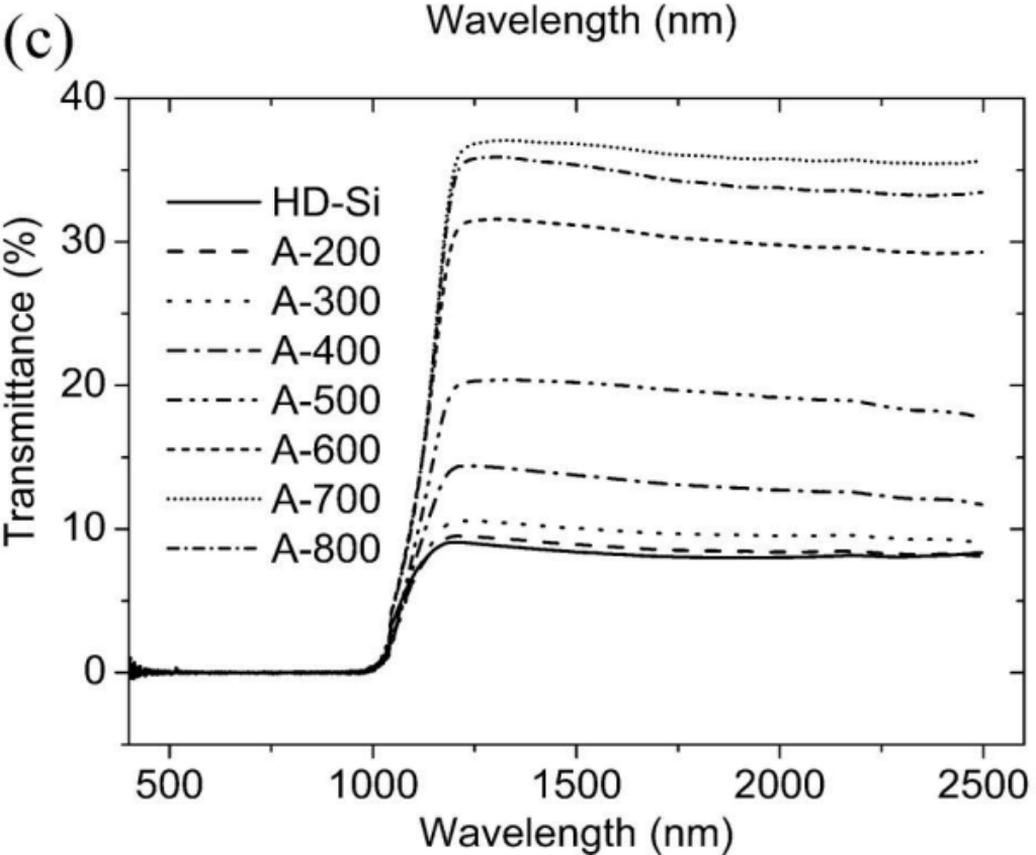

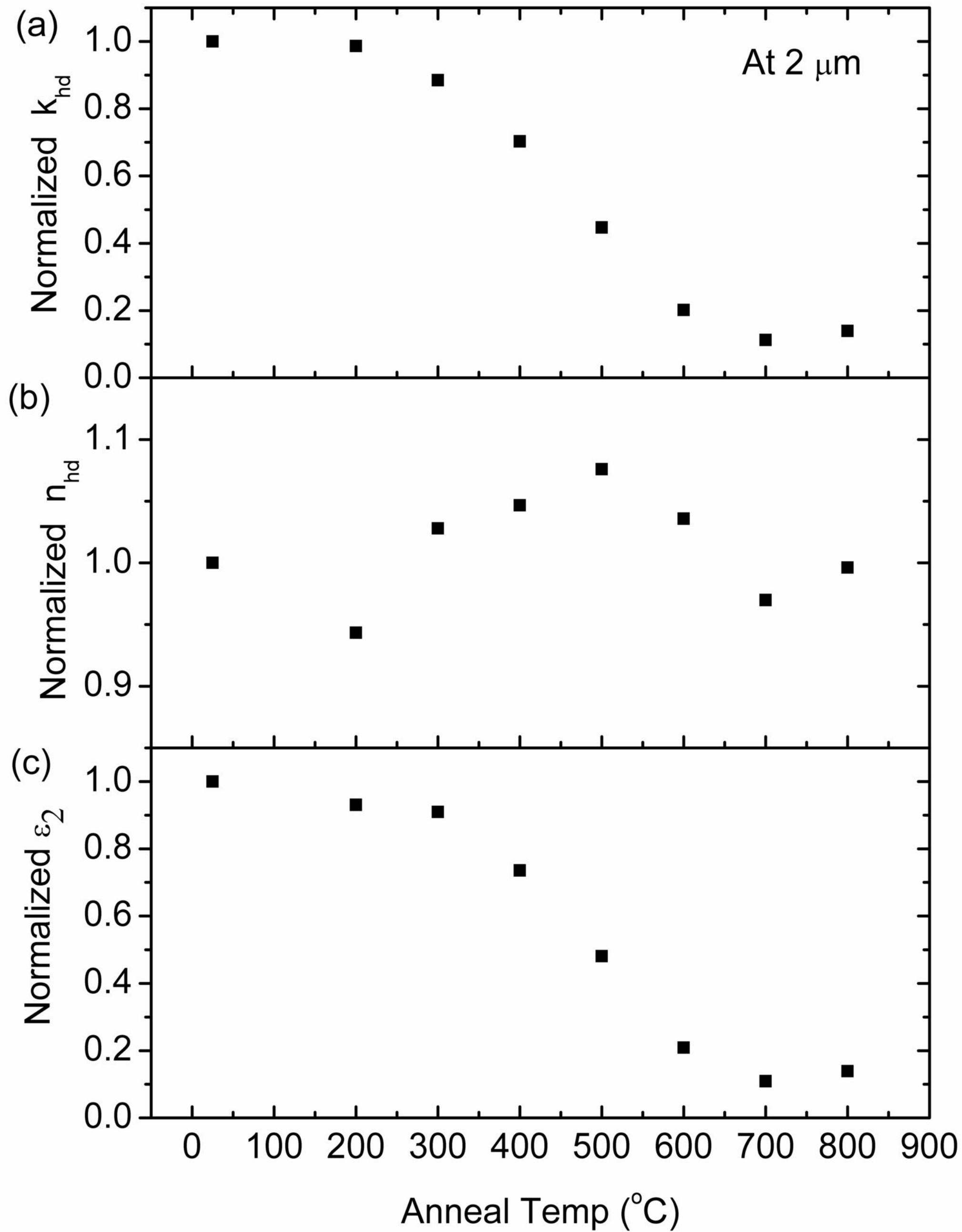

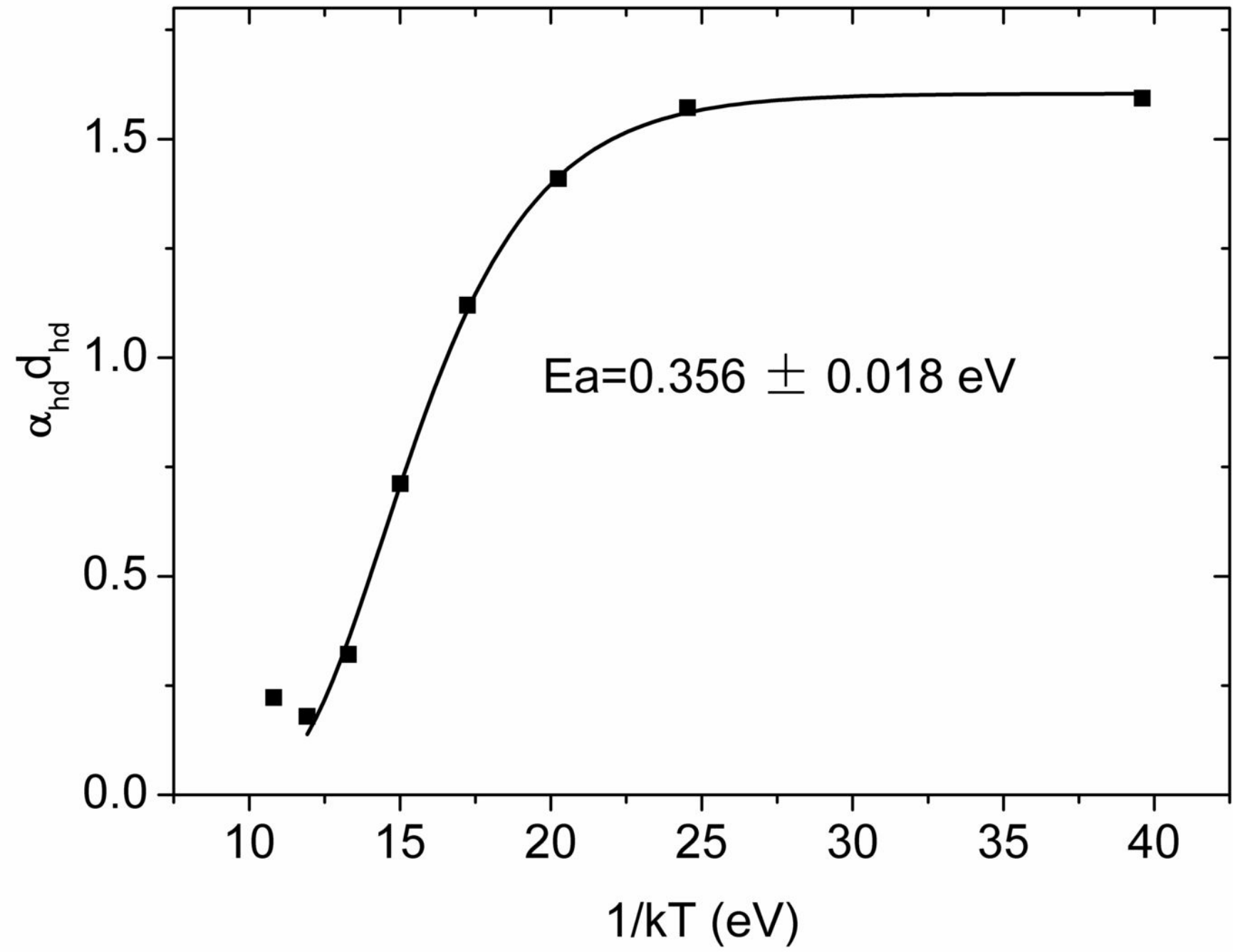

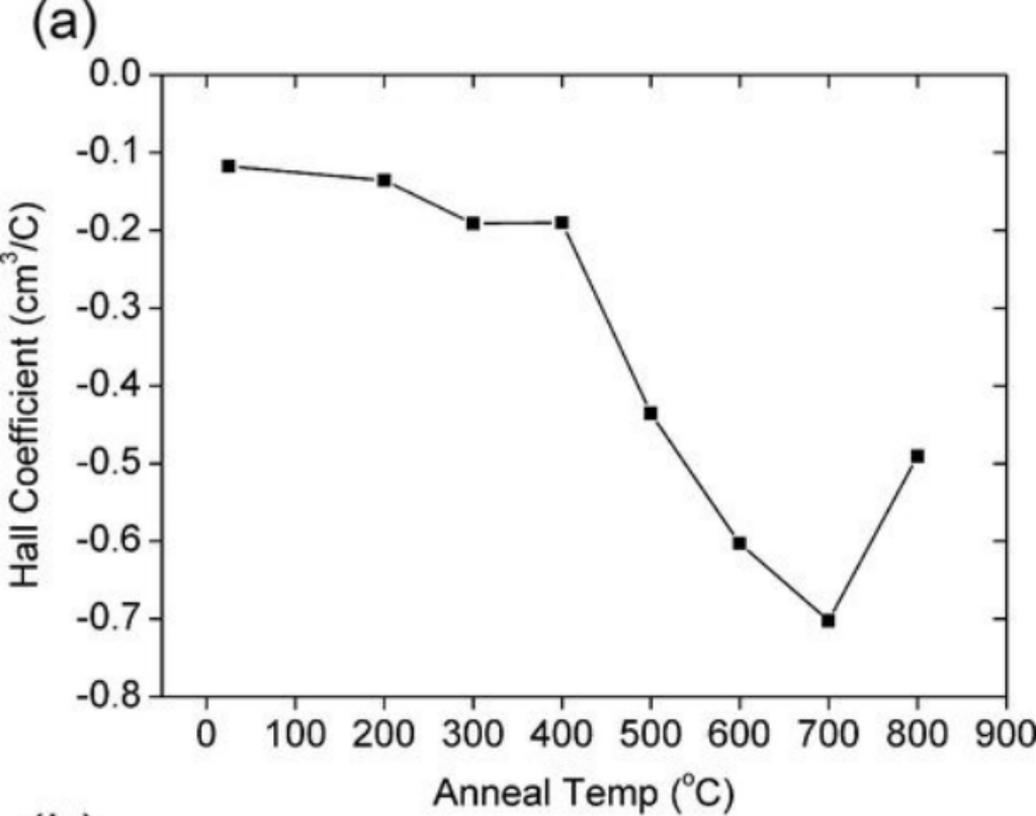

(a)

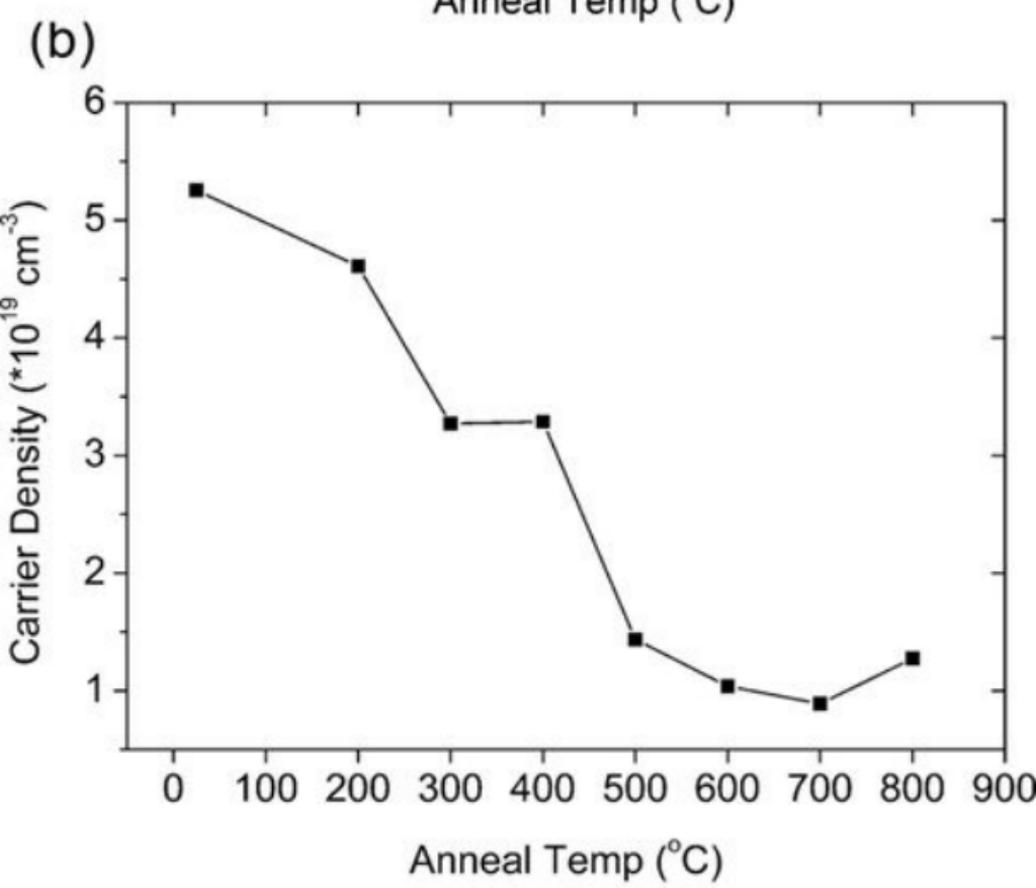

(b)

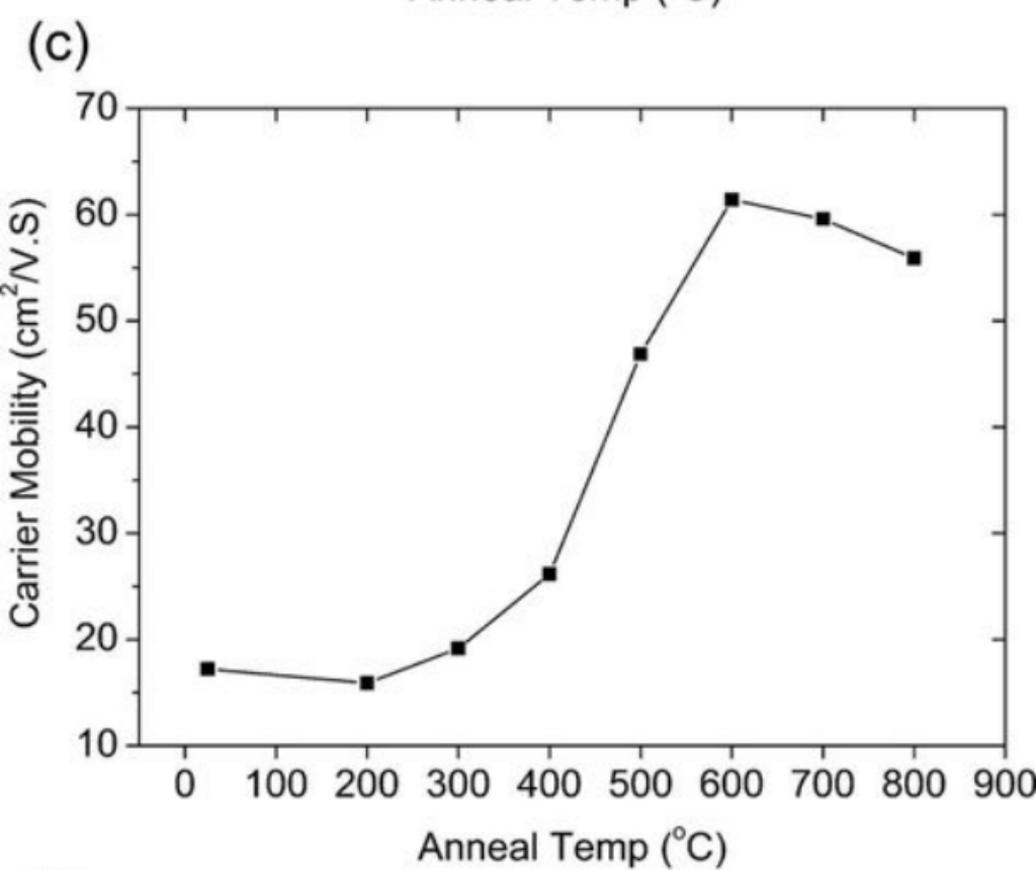

(c)

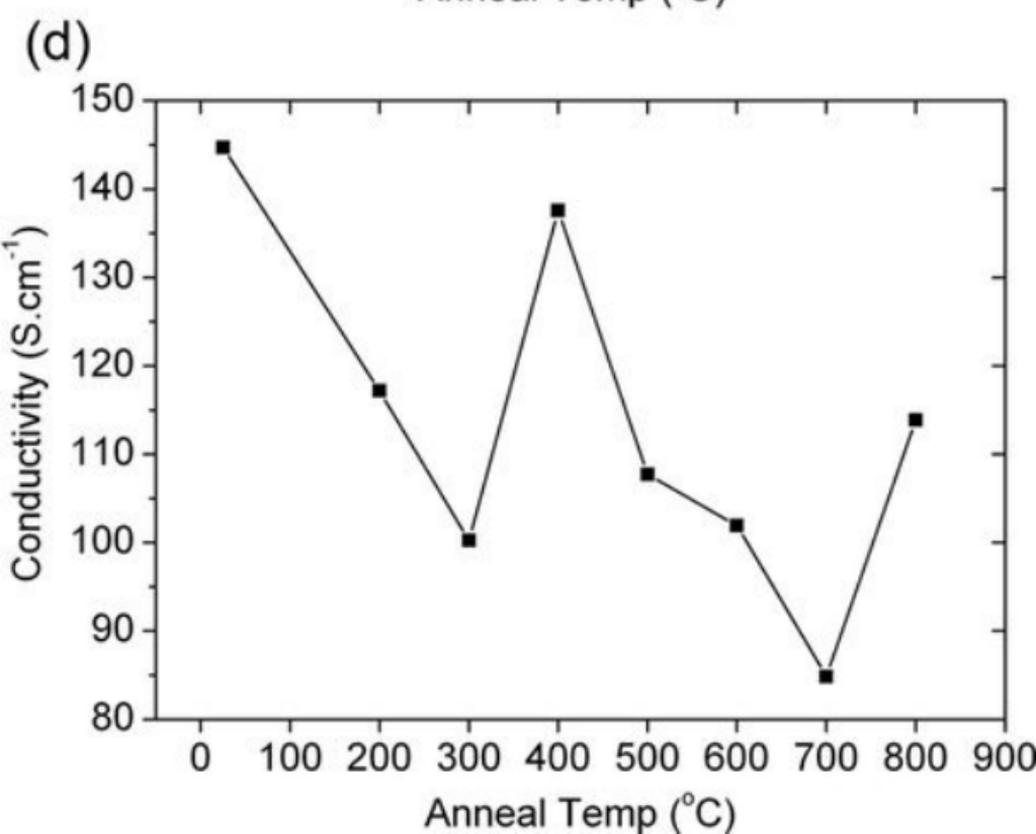

(d)